\documentclass[a4paper]{llncs}

\usepackage{graphicx}
\usepackage{subfigure}
\usepackage[bookmarks=true]{hyperref}

\begin{document}

\title{Essentials of an Integrated\\Crowd Management Support System\\Based on Collective Artificial Intelligence}

\author{Giuseppe Vizzari\inst{1} \and Stefania Bandini\inst{1,2}}
\authorrunning{G. Vizzari, S. Bandini}

\institute{
Complex Systems and Artificial Intelligence Research Centre,\\ University of Milano-Bicocca\\
\email{\{giuseppe.vizzari, stefania.bandini\}@disco.unimib.it}
\and
Researche Center for Advanced Science and Technology,\\
The University of Tokyo
}

\maketitle

\begin{abstract}
The simulation of the dynamical behavior of pedestrians and crowds in spatial structures is a consolidated research and application context that still presents challenges for researchers in different fields and disciplines. Despite currently available commercial systems for this kind of simulation are growingly employed by designers and planners for the evaluation of alternative solutions, this class of systems is generally not integrated with existing monitoring and control infrastructures, usually employed by crowd managers and field operators for security reasons.

This paper introduces the essentials and the related computational framework of an Integrated Crowd Management Support System based on a Collective Artificial Intelligence approach encompassing (i) interfaces from and to monitored and controlled environments (respectively, sensors and actuators), (ii) a set of software tools supporting the analysis of pedestrians and crowd phenomena taking place in the environment to feed a (iii) faster than real-time simulation of the plausible evolution of the current situation in order to support forms of inference providing decision support to crowd managers, potentially directly controlling elements of the environment (e.g. blocking turnstiles, escalators), communicating orders to operators on the field or trying to influence the pedestrians by means of dynamic signage or audible messages.


\end{abstract}

\section{Introduction}

The simulation of the movement of pedestrians and crowds in spatial structures is a consolidated research and application context that still presents challenges for researchers in different fields and disciplines: both the automated analysis and the synthesis of pedestrian and crowd behaviour, as well as attempts to integrate these complementary and activities~\cite{DBLP:journals/expert/VizzariB13}, present open challenges as well as significant opportunities in a smart environment perspective~\cite{DBLP:conf/IEEEcit/SassiBGMMZ15}. Although the currently available commercial tools are used on a day-to-day basis by designers and planners, there is still room for innovations in models, to improve their effectiveness in modeling pedestrians and crowd phenomena, their expressiveness (i.e. simplifying the modeling activity or introducing the possibility of representing phenomena that were still not considered by existing approaches) and efficiency. Moreover, and this is the particular focus of this paper, they are generally not integrated with existing monitoring and control infrastructures, that are employed by crowd managers and field operators.

As testified by an analysis of a recent disaster related to a crowding situation~\cite{Helbing2012}, a more systemic perspective on the overall work flow leading from planning and preparation, to monitoring and control of a crowded event, can help preventing these kinds of situations. This idea of a more thorough and comprehensive computational support to crowd management is also discussed in~\cite{Wijermans2016142}. This paper introduces the notion of an Integrated Crowd Management Support System, a system encompassing both interfaces from and to a monitored and controlled environment (respectively sensors and actuators) as well as a set of software tools supporting the analysis of phenomena taking place in the environment, a faster than real-time simulation of the plausible evolution of the current situation in order to support forms of inference providing decision support to crowd managers, potentially directly controlling elements of the environment (e.g. blocking turnstiles, escalators), communicating orders to operators on the field or trying to influence the pedestrians by means of dynamic signage or audible messages. The following Section provides a brief discussion of the relevant state of the art in pedestrians and crowd automated analysis and synthesis, then some early works providing relevant examples of prototypes of systems partly implementing a crowd control center will be presented and discussed in order to sketch a general architecture. Conclusions and future works end the paper.

\section{Pedestrians and Crowd Analysis and Synthesis}

The research on pedestrian dynamics is basically growing on two lines. On the \emph{analysis} side, literature is producing methods for an automatic extraction of pedestrian trajectories (e.g.~\cite{Boltes2011,DBLP:journals/ijon/BoltesS13}), charactertization of pedestrian flows (e.g.~\cite{DBLP:journals/ijon/KhanBBV16}) automatic recognition of pedestrian groups~\cite{DBLP:journals/pami/SoleraCC16}, recently gaining importance due to differences in trajectories, walking speeds and space utilization~\cite{DBLP:journals/prl/BandiniGV14}. The \emph{synthesis} side -- where the contributions of this work are concentrated -- has been even more prolific, starting from preliminary studies and assumptions provided by~\cite{Henderson1971} or~\cite{helbing1998fluid} and leading to quite complex, yet not usually validated, models exploring components like panic~\cite{DBLP:journals/aamas/BosseHKTWW13} or other emotional variables.

Most of the literature has been focussed on the reproduction of the physics of the system, so on the lowest level, where a significant knowledge on the fundamental diagram achieved with different set of experiments and in different environment settings (see, e.g,~\cite{Zhang2011,Zhang2012}) allows a robust validation of the models. 

Literature of this level can be classified regarding the scope of the modeling approach. Macroscopic models describe the earliest approach to pedestrian modeling, based on analogies between behavior of dense crowds and kinetic gas~\cite{Henderson1971} or fluids~\cite{helbing1998fluid}, but essentially abstracting the concept of individual. A microscopic approach is instead focused on modeling the individual behavior, effectively improving the simulations precision also in low density situations. 

The microscopic approach is as well categorized in two classes describing the representation of space and movement: continuous models simulate the dynamics by means of a force-based approach, which finds its basis on the well-known social force model by~\cite{Helbing1995}. These models design pedestrians as particles moved by virtual forces, that drive them towards their destination and let them avoid obstacles or other pedestrians. Latest models in this class are the centrifugal force model by~\cite{Chraibi2010} and the stride length adaptation model by~\cite{VonSivers2014}. Other examples consider also groups of pedestrians by means of attractive forces among persons inside the group~\cite{Qiu2010a}.

The usage of a discrete environment is mostly employed by the cellular automata (CA) based models, and describes a less precise approach in the reproduction of individuals trajectories that, on the other side, is significantly more efficient and still able to reproduce realistic aggregated data. This class derives from vehicular modeling and some models are direct adaptations of traffic ones, describing the dynamics with ad hoc rules (e.g.~\cite{Blue2000,BlueAdlerCACrowd}). Other models employs the well-know \emph{floor field} approach from~\cite{Burstedde2001}, where a \emph{static} floor field drives pedestrians towards a destination and a \emph{dynamic} floor field is used to generate a lane formation effect in bi-directional flow.~\cite{Suma2012} is an extension of the floor field model, introducing the \emph{anticipation} floor field used to manage crossing trajectories and encourage the lane formation.~\cite{Kirchner2004} discussed methods to deal with different speeds, in addition to the usage of a finer grid discretization that decreases the error in the reproduction of the environment, but significantly impacts on the efficiency of the model. An alternative approach to represent different speeds in a discrete space is given by~\cite{CrocianiTGF}. ~\cite{CASM2013} is another extension of the floor-field model, where groups of pedestrians are also considered. 

\begin{figure}[t]
\begin{center}
\includegraphics[width =.85\textwidth]{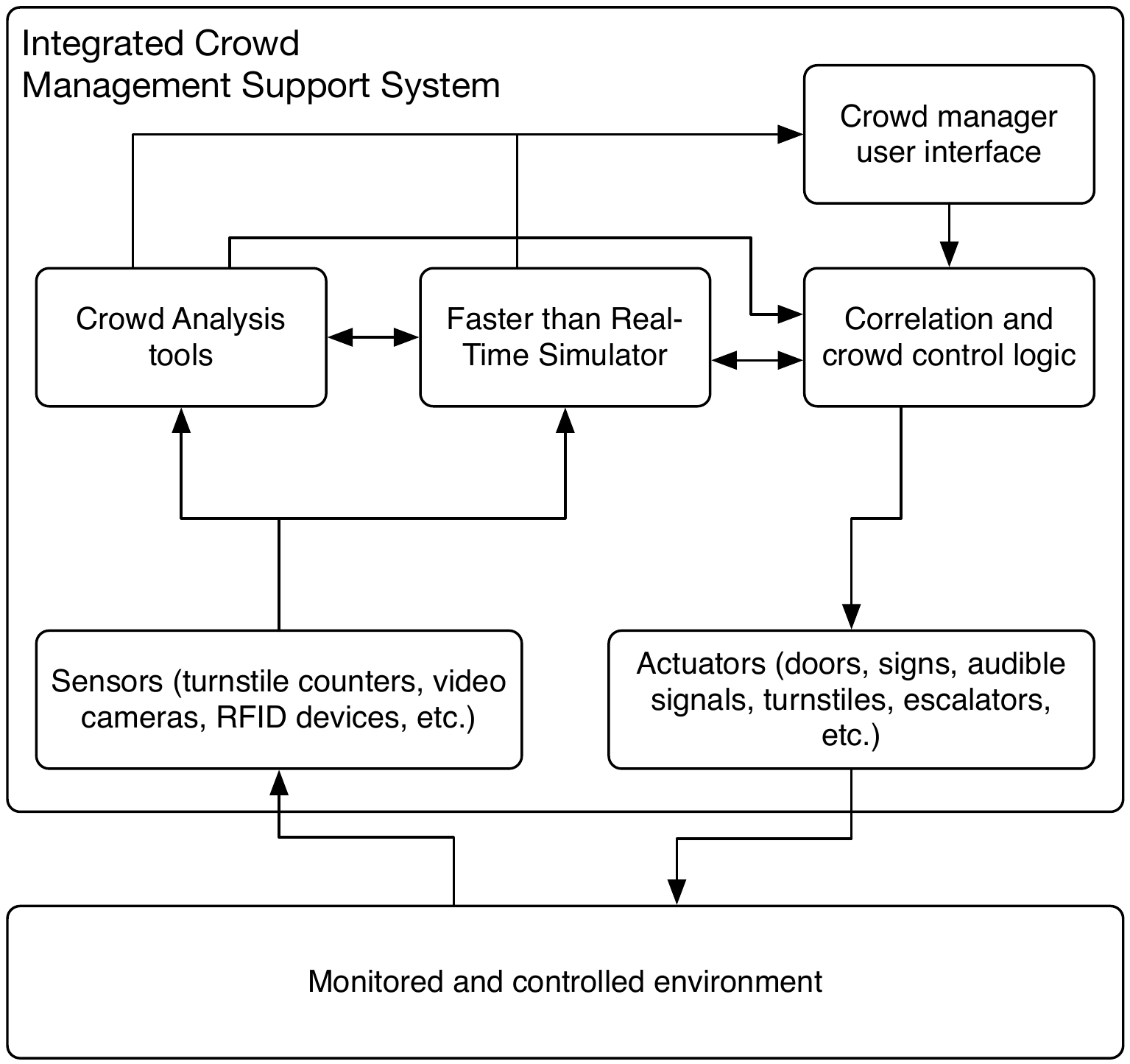}
\caption{A high level architecture of an Integrated Crowd Management Support System.}\label{fig:arch}
\end{center}
\end{figure}

\section{Towards an Integrated Crowd Management Support System}

The first work proposing the idea of employing faster than real-time pedestrian simulation to support decision making activities of crowd managers was proposed in~\cite{DBLP:journals/sj/GeorgoudasSA11}: the authors were essentially considering the idea of feeding a CA-based simulator with data from computer vision algorithms characterizing a given evacuation situation. The simulation of a short-term future could detect plausible congested positions and, therefore, actions influencing pedestrian movement (i.e. activation of sound and/or optical signals) could be enacted to prevent them. As for a more recent work by the same research group~\cite{DBLP:journals/sj/TsiftsisGS16}, the authors also describe a hardware (FPGA) implementation of the system, for achieving increased performance of both the computer vision and CA simulation systems.

Another relevant prototypical implementation of this kind of system was carried out in the context of the Hermes project~\cite{DBLP:journals/aes/WagoumSSC13,DBLP:journals/jvca/WagoumCMSS12}: once again, also this work basically considered evacuation situations, in which the intention of pedestrians is generally to vacate the area in the shortest possible time, which generally means following the shortest path.

Both the above mentioned works present separate blocks for the interpretation of inputs to the system (mostly computer vision techniques for analyzing video footage and perform pedestrian detection, counting, tracking) and the short-term simulation of the plausible evolution of the system. Both works simply suggest, but do not actually describe in details, the presence of a module actually coordinating these modules: triggering the start of a simulation according to the passage of time (i.e. repeatedly perform a 10 minute simulation every minute) or due to the detection of a certain kind of event (e.g. an increased incoming flow from a certain entrance), interpreting simulation results (e.g. to detect and characterize congested areas), correlating these results and available information on the spatial structure of the environment (e.g. regions and passages connecting them) to support decision making activities by users, that are generally supposed to be ``in--the--loop'' (i.e. actually taking final decisions on actions to be enacted). The overall resulting architecture can be described by the schema depicted in Figure~\ref{fig:arch}: an Integrated Crowd Management Support System is essentially responsible to support the control of an area subject to crowding conditions calling for a careful planning of the operations, monitoring of actual crowd dynamics, decision support to the crowd manager (and potentially supporting a smooth communication with field operators) and potential enactment of actions (e.g. different forms of information provisioning to pedestrians or actual changes in the environmental conditions) to influence overall crowd dynamics.  

A different and additional line of work, described in~\cite{Lujak2016} and~\cite{OssowskiATT16}, and developed in the context of a Spanish research project, does not employ pedestrian simulation systems, but it rather relies on consolidated results in the area of route optimization to provide guidance in smart cities, large public events and evacuation situations. This work is in earlier stages compared to the previously described ones, but it is interesting to mention both to suggest that predictions on system evolution can be carried out also by more coarse grained models of the environment, and also to stress the fact that a growing number of additional sensors can be considered to complement and enrich the information acquired by means of computer vision tools. These works also suggest that different Crowd Management Support Systems could be fruitfully connected, both to leverage the monitoring capabilities and be able to foresee changes in the local situations whenever the monitored areas are connected and reachable from one another, and more generally to support a more comprehensive and systemic view of the overall environment, reducing at the same time the computational costs related to the simulation of a potentially very high number of pedestrians. 

Finally, this line of work suggests that, besides supporting decisions of experts, either designers or crowd managers, this kind of system could be used to provide a distributed awareness of the state of the overall (large scale) environment to actually trigger \emph{collaborative} forms of diffused, bottom-up behaviours that, at the same time, lead to individually and collectively beneficial choices: by being informed about the state of a transportation network and informing (either explicitly or implicitly, through a behaviour implicit form of communication) the overall system, pedestrians could avoid situations of congestion (causing delays or even simply unpleasant travels), making a reasonable individual choice that could also play a role in limiting the growth of the already existing congestion. Although research on this line of work in the transportation area is still quite open~\cite{DBLP:journals/scpe/Varga15}, the idea of investigating a \emph{collective intelligence} scheme~\cite{AI-and-CI} in which different forms of AI techniques are employed is appealing and promising.

\section{Conclusions and Future Developments}

This paper has introduced the notion of an Integrated Crowd Management Support System, an comprehensive set of tools supporting the both analysis of crowd dynamics and its simulation (faster than real-time) for supporting short term predictions on the evolution of the monitored environment. A control logic module is responsible both to trigger the simulation, to analyze its results and supply the crowd manager indications on potential issues and potential mitigation actions. Whereas the state of the art presents abundant and significant results both in the analysis of crowd dynamics and in their simulation (even respecting the faster than real-time speed requirement) more work has to be carried out to define a proper control logic for a Crowd Management Support System: this module should encompass both the ability to analyze and correlate the flow of information coming from sensors, to trigger simulation and other forms of predictions, but also to support decisions on the most appropriate mitigation actions to prevent or solve potential issues.

\section*{Acknowledgements}
The authors would like to thank Davide Bogni, Guido Bonazza, Luca Crociani, Andrea Gorrini and Sara Manzoni for the fruitful discussions that contributed to the conceptualization of the Crowd Management Support System concept and architecture.


\end{document}